\documentstyle[prb,preprint,aps]{revtex}
\begin{document}
\draft
\title{Vibrational Spectra of Defects in Silicon:\\ An Orbital Radii
Approach}
\author{H.C.Verma\thanks{On leave from Science College,
Patna-800005,INDIA},  George C. John, and Vijay A. Singh}
\address{Physics Department, I.I.T.-Kanpur, U.P. 208106, INDIA}
\date{May 1995}
\maketitle
\bibliographystyle{prsty}
\begin{abstract}
	A  phenomenological     approach to the    stretching mode
vibrational   frequencies   of  defects    in   semiconductors  is
proposed. A novel  quantum scale is defined in  terms of the first
principles  pseudopotential based  orbital  radius r$_s$  and  the
principal  quantum number of the  element  concerned.  A universal
linear relationship  between the Sanderson electronegativity ({\it
SR} ) and this quantum  scale is established.   Next, we show that
the   stretching  mode   vibrational   frequencies   of   hydrogen
($\nu_{Si-H}$) and chlorine ($\nu_{Si-Cl}$) in the silicon network
scale    linearly with this     quantum  scale.  Predictions   and
identifications  of defect environments  around the Si-H and Si-Cl
are   possible.  The  assignments of  vibrational  modes in porous
silicon are critically examined. We  discuss our proposed scale in
the context of    Mendeleveyan  scales in general,   and   suggest
justifications  for  it.  We  believe   that our  approach  can be
gainfully   extended to     the   vibrational spectra   of   other
semiconductors.
\end{abstract}

\section{Introduction}
	The vibrational spectra of  defects in silicon lie in  the
infrared  and   the  far  infrared  region.   They   constitute an
important signature,   enabling  one  to identify  the  particular
defect  and  its immediate  environment. There has  been a renewed
interest  in their  study with   the recent  discovery  of visible
photoluminescence    (PL)   in   porous   silicon   \cite{canh90}.
Explanations  for  this    noteworthy    phenomenon are     varied
\cite{john95r}. Suggestions invoking  molecular  complexes such as
siloxene  \cite{bran92}, Si-H$_2$ \cite{tsai92} and   non-bridging
oxygen-hole centers \cite{prok94} are based, at  least in part, on
comparing  their vibrational frequencies   with those  observed in
porous silicon.   It has also  been  suggested that the Si-H  bond
plays   an important  role in   visible  PL, since  porous silicon
becomes non-radiative with decreasing H content \cite{ohno92}.

	It has  been known  for  sometime  that  the stretching  mode
vibrational frequencies of Si-H ($\nu_{Si-H}$) in  substituted silane
molecules SiHR$_1$R$_2$R$_3$ \cite{smit59},  amorphous solids such as
a-Si  and  a-SiO$_2$  \cite{luco79},  and   in  crystalline   silicon
(c-Si)\cite{shi82}   correlate  with   the   electronegativities   of
\{R$_1$R$_2$R$_3$\}.  The following explanation is usually proffered
for  this correlation. As  the  electronegativity of the substituting
species \{R$_1$R$_2$R$_3$\}  increases,  the s-character of  the Si-H
bond increases.  A  calculation of the s-component  of the bond-order
matrix  substantiates  this  \cite{sahu82}.Because  of  the  enhanced
s-character, the Si-H distance $d_{Si-H}$ decreases and the effective
force-constant increases.  Thus $\nu_{Si-H}$ increases.

	The  Sanderson (and not  the  Pauling)  electronegativity  is
employed in explaining the trends in $\nu_{Si-H}$ \cite{luco79}.  The
former is known to  correlate  well with the structural properties of
molecules and  is also  called the stability-ratio  electronegativity
{\it  SR}.   The  importance  of  classical  and  quantum  scales  in
solid-state phenomenology can hardly be overemphasized \cite{phil78}.
We propose in this work  to  employ  a quantum  scale, based  on  the
orbital   radii    generated   by   first   principles   calculations
\cite{zung80},  and  the  principal  quantum  number  to  systematize
$\nu_{Si-H}$ and the chlorine related  frequencies $\nu_{Si-Cl}$. The
advantages of this scale are: (i) it is derived from first principles
calculations  on free atoms.   In  other  words it  is non-empirical,
fixed, and  not  subject to periodic  updating  unlike the  empirical
scales  like  electronegativity; (ii) it is  defined for all elements
(except  H).  For  example  it is defined for the group VIIA elements
 \{F,Cl,Br,I\} unlike the semi-empirical Miedema scales \cite{mied73}.
The  orbital radii  have  been  gainfully  used  to  systematize  the
structure  of  binary  alloys  \cite{zung80},  the  phenomenology  of
ion-implantation sites \cite{sing82},  and more recently to construct
quantum   structural   diagrams   for   high   T$_c$  superconductors
\cite{vill88}  as well as explain trends in binding energies obtained
by local density approximation (LDA)\cite{yeh92}.

	In section  II, we introduce a  novel quantum scale ${\cal
V}$ based on the principal  quantum number and the orbital  radius
$r_s$.  We   call  it the  valence  shell \underline{re}normalized
\underline{el}ectronegativity (REEL).   We demonstrate  that    it
correlates linearly with   the  Sanderson  electronegativity   and
further, with the stretching  mode vibrational frequencies of H in
Si. We describe the utility of  our scale in assigning vibrational
modes  to impurities  (impurity complexes)  in  porous silicon. In
section III,  we  discuss   our scale   in the context   of extant
phenomenological scales and seek to justify it qualitatively.

\section{Vibrational Spectra of Defects in Silicon}
\subsection{Electronegativity and Orbital Radius}

	To motivate the application of the orbital-radii scales to
the systematization and prediction  of the vibrational frequencies
of H and other impurities in semiconductors, we shall first relate
the Sanderson electronegativity $SR(R_j)$ to the orbital radius of
the  element  $R_j$. The  three  orbital radii  listed  by  Zunger
\cite{zung80}  are r$_s$, r$_p$,  and r$_d$ which are respectively
the  crossing points of the  first principles pseudopotentials for
$l$ = 0, 1, and 2. The role of the $s$ character of a $\nu_{Si-H}$
bond in  determining  the  Si-H vibrational frequencies   has been
pointed   out by  other  workers (refs.  9 and   10 in the present
work).   Further, $\nu_{Si-H}$  has   been  related to   Sanderson
electronegativity   {\it SR}  of   element  $R_j$  (ref.8).   This
motivated us  to seek a  relation between  the  $SR(R_j)$ and  the
Zunger orbital radii $r_s (R_j)$.

  Since the s-character of the Si-H bond appears to play a key role
\cite{sahu82}, we attempt  to relate  $SR(R_j)$ to $r_s(R_j)$.  It is
known  that  the  interatomic   distance  decreases  with  increasing
electronegativity  difference. Further, the orbital radius $r_s(R_j)$
demarcates the region between  the  inner  repulsive Pauli  potential
($<r_s(R_j)$)   and   the   outer   attractive   Coulomb    potential
($>r_s(R_j)$). The  smaller $r_s(R_j)$  is, the larger is the spatial
extent of the attractive Coulomb and exchange  potentials.  For these
reasons, we seek a relationship of the form
\begin{eqnarray}
SR(R_j) \sim \frac{1}{r_s(R_j)} \label{eq:1} 
\end{eqnarray}

	Such an inverse relationship between the electronegativity
and     orbital    radii      has      been     posited    earlier
\cite{john74,chel78,burd82,zung78}.   We       shall   discuss the
relationship of our work to previous works  in section III. It was
found    that  a  linear  relationship  of   the    form  given by
eqn.~(\ref{eq:1}) can be established for each  row of the periodic
table. This is depicted in figure  1. But to establish a universal
linear relationship  incorporating several  rows of   the periodic
table,  we need   to postulate  another  quantum  scale.  In  this
connection, we recall that the structural separation plots for the
binary  alloys at   times  employ the   principal  quantum numbers
$n(R_j)$ of  the constituent   elements.  The Mooser-Pearson  plot
uses  the two   scales: electronegativity  and   principal quantum
number  $n$  of  the  valence  shell\cite{pear69}.   The Shaw plot
employs electronegativity and the   cube of the  principal quantum
number   \cite{shaw68}.  According  to   Zunger  \cite{zung80} the
orbital  radii  are characteristic   of   the atomic cores   whose
defining quantum numbers   are $1,2,...,n-1$.  In  this spirit  we
attempt  to  correlate  the  electronegativity $SR(R_j)$  with the
functional  form  $f((n(R_j)-1),r_s)$.  An  inspection of figure 1
reveals that there is  a systematic shift in the  slope of the row
wise plots. This insight, and some trial and  error, led us to the
discovery that {\it SR} scales linearly with the quantity
\begin{eqnarray}
{\cal V}(R_j)= \frac{\sqrt{n(R_j)-1}}{r_s(R_j)} \label{eq:verma}
\end{eqnarray}
which we call the valence shell \underline{re}normalized
\underline{el}ectronegativity (REEL). 

 Fig.2 depicts a linear scaling behavior:
\begin{eqnarray}
 SR(R_j) = a {\cal V}(R_j)+ S_0    \label{eq:2}
\end{eqnarray}

Here the slope $a$ and  the intercept $S_0$  are 2.07 ($\pm$ 0.02)
(a.u.)$^{-1}$  and   -1.66  ($\pm$  0.006)    respectively.    The
uncertainty in the calculated electronegativity  is $\pm$ 0.01 and
is  of  the  same  order   as  the  uncertainty in   the  original
electronegativity data  \cite{sand60}.  For larger values there is
a tendency to saturation. Our attempts to correlate $SR(R_j)$ with
the  orbital radius r$_p$ or ($\mbox{r}_s  + \mbox{r}_p$) were not
as successful.

\subsection{Vibrational Spectra}

	We now establish the relationship between the Si-H stretching
frequencies  $\nu_{Si-H}$ observed in  silicon  and the quantum scale
defined by  us. Table  I lists some of the well known Si-H stretching
bands observed in crystalline  silicon  (c-Si) and  amorphous silicon
(a-Si).  The frequencies listed in the first  seven rows are observed
for c-Si and the next  four for a-Si. The last three  entries involve
dangling bonds  and are observed  in  c-Si. The values are taken from
the data culled by Shi et al.\cite {shi82}  and Kniffler et al. \cite
{knif83}.   As   demonstrated   by  fig.  3,  one  obtains  a  linear
relationship between $\nu_{Si-H}$ and our quantum scale.
\begin{eqnarray}
\nu_{Si-H} = m {\cal V}(R_j) + \nu_0  \label{eq:3}
\end{eqnarray}
Here the summation is  over   all four  nearest neighbors of   the
central Si atom  and includes H. For H  we have taken the relevant
REEL  value \mbox{$\cal  V$}  to be  2.86.   For the dangling bond
neither a $n(R_j)$ nor a $r_s(R_j)$ can  be defined. To include it
into our scheme  we take \mbox{$\cal V$}  to be 0.54. These values
have been obtained from empirical fit to data. In eqn.~\ref{eq:3},
the  slope $ m  = 61.03 $cm$^{-1}$-(a.u.)$^{-1}$ and the intercept
$\nu_0 =  1456.45 $cm$^{-1}$. The  uncertainties in  the slope and
the intercept  are $\Delta m  = \pm  0.07$ cm$^{-1}$-(a.u.)$^{-1}$
and  $\Delta   \nu_0    =\pm7.3   $cm$^{-1}$   respectively.   The
relationship given by eqn.~\ref{eq:3}   has predictive value  with
the  attendant uncertainty $\Delta \nu   = \pm (11-16) $cm$^{-1}$.
We shall  employ  it  to  obtain Si-H  stretching  frequencies  in
environments likely to occur in porous silicon.

	So  far we  have  discussed  Si-H  vibration  frequency  on a
lattice  or an amorphous network. One may  also extend  this work  to
$\nu_{Si-H}$ in silane molecules substituted by organic radicals such
as CH$_3$, C$_2$H$_5$, C$_6$H$_5$ etc. The {\it SR}  values for these
radicals  can  be  calculated  using  Lucovsky's  prescription  \cite
{luco79} (his eqn.(2)). The linear fit in fig. 2 enables us to define
an effective \mbox{${\cal V}(R_j)$} using these {\it SR}  values. One
may then look for a correlation between $\nu_{Si-H}$ and these values
of \mbox{${\cal V}(R_j)$}. We  have carried out such an exercise  for
SiH({\it R}Cl$_n$) (n = 0,1,2) where {\it R} stands  for the radicals
and found an approximate linear relationship akin to fig.3.

	Chlorine is known  to be  a good dangling bond passivator  in
silicon. We have also  discovered a linear  relationship between  the
Si-Cl  stretching mode  frequency and our quantum scale. We  use  the
data for $\nu_{Si-Cl}$ in a-Si cited by Wu et al. \cite{wu83}. Fig. 4
depicts  this  correlation.  Except  for  the  highly electronegative
environment (Cl$_3$)SiCl, the data is linear to a good approximation.

\subsection{Porous Silicon}
	Infrared  frequencies  observed in  porous silicon  in the
range  2050-2150 cm$^{-1}$ have  been attributed to the stretching
mode of the Si-H bond. Specifically,  three broad peaks (halfwidth
$\approx 20 $cm$^{-1}$) are observed in this range which have been
attributed   in  the past  to  the  nearest  neighbor environments
(Si$_3$)SiH,           (Si$_2$)SiH$_2$        and      (Si)SiH$_3$
\cite{unag80a,unag80b,gupt88,kato88,koch93b}.     However,    some
ambiguities still remain   about the detailed assignments   of the
stretching  mode vibrations.  Gupta  and  co-workers \cite{gupt88}
have shown that during thermal annealing,  both the 2087 cm$^{-1}$
and the  910  cm$^{-1}$  peak disappear simultaneously.    The 910
cm$^{-1}$   peak has been traditionally   assigned to the Si-H$_2$
scissors mode.  Therefore,  they have assigned  the 2087 cm$^{-1}$
line to the Si-H$_2$  stretching mode and  the 2110 cm$^{-1}$ line
to the Si-H  stretching  mode.  Other workers   \cite{kato88} have
assigned the 2090, 2110 and  2140 cm$^{-1}$ line to Si-H, Si-H$_2$
and Si-H$_3$ stretching frequencies respectively.

	 The present  study shows  that the vibrational  frequency
increases with increasing hydrogen content (eqn. 4).  This is also
supported   by theoretical calculations  based  on {\it ab initio}
molecular  orbital  approaches \cite{ogat95}.  The Si-H stretching
frequency is reported \cite{shi82} to   be $\sim$ 2000  cm$^{-1}$.
The  frequency for    Si-H$_3$ stretching   mode calculated   from
equation \ref{eq:3} is $\sim$ 2100 cm$^{-1}$ (see  table 2).  This
suggests the   assignment of 2087 cm$^{-1}$   to Si-H$_2$ and 2110
cm$^{-1}$ to Si-H$_3$ stretching     modes, which is seen  to   be
consistent with  the  experimental results  of Gupta,  Colvin  and
George and the theoretical results of the  present work as well as
Ogata and  co-workers  \cite{ogat95}.  The 2140 cm$^{-1}$ line can
be assigned to an oxygen  complex (OSi)Si-H$_2$ as calculated from
equation \ref{eq:3}.  A  similar  suggestion  has also  been  made
elsewhere in literature \cite{gupt88}.

	An interesting relationship between the Si-H bond distance
$d_{Si-H}$          and      $\nu_{Si-H}$     exists,       namely
$d_{Si-H}^3$$\nu_{Si-H}$ =   7074 \AA$^3$cm$^{-1}$.  The  range of
$\nu_{Si-H}$ observed  in porous silicon  suggests that $d_{Si-H}$
is confined to \{1.49 - 1.51\}\AA.  This should  be a useful guide
in electronic structure  calculations where the dangling bonds are
passivated by hydrogen.  Values such  as $d_{Si-H}$ = 1.637 \AA or
$d_{Si-H} =    1.17    $\AA   employed in    these    calculations
\cite{read92,sand92} are clearly out of range.

\section{Discussion}

	Phenomenological scales have  been employed to systematize
large databases  in condensed  matter physics for  quite sometime.
The  phenomena  studied  include crystal  structure of  binary and
ternary alloys, solid  solubilities, heats of formation of alloys,
locations of  ion implantation  sites and  structural  diagrams of
high $T_c$  superconductors,  among  others.  The oldest  pair  of
scales,  the   principal  quantum number  $n(A)$   and the valence
electron  number  $Z_v(A)$ of the element   A form the  basis of the
periodic table   and   in  this spirit,  phenomenological   scales
discussed   herein  are referred to  as   Mendeleveyan scales. The
Darken-Gurry plots \cite{dark53,wabe63} are based  on the pair  of
scales  \{ $r(A),  \chi(A)$ \}  where $r(A)$ is  the atomic radius
(usually, the   Goldschmidt radius) and  $\chi(A)$ is  the Pauling
electronegativity of the element  A. The principal  quantum number
$n(A)$ and  the Pauling electronegativity $\chi(A)$ constitute the
Mooser-Pearson scales  \cite{pear69}.  The Shaw plot \cite{shaw68}
as   mentioned in Sec. II  employs  $\chi(A)$ and  the cube of the
principal quantum number $n(A)$.  The Hume-Rothery rules for alloy
structures   cite   the atomic size    mismatch   and the electron
concentration per atom as the  relevant variables.  To systematize
the data  on heats of formation  of alloys, Miedema  and co-workers
\cite{mied73} employed purely  quantum  scales,  namely, the  work
function $\phi(A)$ (which can  be related to the electronegativity
$\chi(A)$ ) and the electron  density at the Wigner-Seitz boundary
$n_{WS}(A)$.   Refinements  of   the  Miedema   scales have   been
suggested by several  workers \cite{chel78,wats78}.  On the  other
hand Bloch and co-workers \cite{john74,simo73} suggested the use of
orbital  radii  \{$r_s(A),r_p(A),r_d(A)$\}  derived from  free ion
quantum defects  to separate crystal   structure of alloys. Zunger
\cite{zung80} defined non-empirical orbital  radii in terms of the
first principles  hard-core pseudopotentials and  employed them to
systematize the  crystal  structures of around 500  binary alloys.
Recently,  Villars  and    Hulliger \cite{vill87}  carried   out a
systematization of  the  ternary  alloys using   three dimensional
plots   based   on  the   orbital radii     ($r_s(A)+r_p(A)$), the
Martynov-Bastanov  electronegativity and  the    valence  electron
number.

	We may classify the  phenomenological scales as follows:\\
(i)\underline{Integer  Scales}:    examples  are     the  original
Mendeleveyan scales,  namely the principal  quantum number $n$ and
the    valence electron  number   $Z_v$.  \\ (ii)\underline{Length
Scales}:   for   example   the  Goldschmidt     radius employed in
Darken-Gurry or   Mooser-Pearson  plots, the Ashcroft   empty core
radius, and the   Pauling tetrahedral and  octahedral  radii. In a
sense,  these denote  the    ``size''  of an  atom.    Anisotropic
l-dependent length scales  are the orbital radii \{$r_s,r_p,r_d$\}
of  Zunger and    of  Bloch  and  co-workers,    the  former  being
non-empirical in character.\\ (iii)\underline{Electronegativity} :
the Pauling electronegativity is commonly employed. Alternatively,
the work function  $\phi$, the chemical potential, the  ionization
potentials, or   the  Sanderson  electronegativity  maybe  used.\\
(iv)\underline{Miscellaneous Quantum  Scales}: among the important
and useful ones  are  the electron   density at the   Wigner-Seitz
boundary $n_{WS}$ (a Miedema scale) the average covalent and ionic
energy  gaps  (Phillips and  Van   Vechten \cite{phil70}) and  the
electron concentration per atom (Hume-Rothery).

	An issue of relevance to the present  study is whether the
scales are  independent, in particular,  the relationship, if any,
between the electronegativity  scale and the Zunger orbital radii,
which     is     a  length    scale.       The  Darken-Gurry  plot
\cite{dark53,wabe63}  treats  the electronegativity and the atomic
size as independent variables. More recently, Villars and Hulliger
\cite{vill87} have considered the electronegativity and the sum of
the orbital radii ($r_s+r_p$)  as independent  variables. However,
there exists a substantial body of studies which posits an inverse
relationship between the Pauling electronegativity and the orbital
radii.  St.  John and Bloch \cite{john74} as well as in a detailed
study  Chelikowsky and Phillips  \cite{chel78} have suggested that
an electronegativity  scale  maybe defined in  terms  of  a linear
combination of
\{$r_s^{-1},r_p^{-1},r_d^{-1}$\}. In other words
\begin{eqnarray} 
\tilde{\chi} = \tilde{\chi_0} + \sum_{l=0}^{l=2} \frac{a_l}{r_l}
\end{eqnarray} 
where $\tilde{\chi}$ is the arithmetic mean of the Pauling and the
Phillips electronegativities. Zunger \cite{zung78} suggests   that
the  Pauling electronegativity may  scale with the inverse orbital
radius.  Burdett  \cite{burd82}  has also  noted that the  Pauling
electronegativity is related to the Zunger orbital radii
\begin{eqnarray} 
\chi = A\left[ \frac{1}{r_s} + \frac{1}{r_p} \right] + \chi_{0B}
\end{eqnarray} 
where $A$ and $\chi_{0B}$ are constants.

	It is of interest to note a study by Watson and Bennett
\cite{wats78} in which the Pauling electronegativity is related to
the s and p ionization potentials $E_s$ and $E_p$ respectively.
They work with the hybridized expression
\begin{eqnarray} 
\chi = 1.075 (E_s + 3E_p) + 0.35
\end{eqnarray} 
However they also note that a linear scaling relationship exists
between $\chi$ and $E_s$ and separately between $\chi$ and $E_p$.

	Our search for a universal relationship between
electronegativity and the Zunger orbital radius had its genesis in
the above-mentioned observations. To motivate our work further we
note that the Pauling electronegativity $\chi(A)$ and the valence
electron number $Z_v(A)$ can be related \cite{chel78}
\begin{eqnarray*}
2 \chi(A) = Z_v(A) + C
\end{eqnarray*}
where C is a constant. The quantum defect radii are related to
$Z_v$ as follows
\begin{eqnarray*}
Z_v = \frac{\hat{l}(\hat{l}+1)}{r_l}
\end{eqnarray*}
where $\hat{l}$ is an $l$-dependent parameter. It follows that 
\begin{eqnarray} 
\chi(A) \propto \frac{1}{r_l}
\end{eqnarray} 

	A similar viewpoint is obtained by extending the work of
Watson and Bennett \cite{wats78}. Following them, 
\begin{eqnarray} 
\chi \propto E_s
\end{eqnarray} 
The ionization potential $E_{n,l}$ is given in quantum defect
theory by
\begin{eqnarray*}
 E_{n,l} = \frac{- Z_v^2}{2(n + \hat{l} - l)^2}
\end{eqnarray*} 
Taylor expanding the above expression
\begin{eqnarray*}
E_{n,l} \simeq \frac{- Z_v^2}{2 \hat{l}^2} +
 \frac{- Z_v^2}{\hat{l}^2}\left(\frac{n-l}{\hat{l}}\right) +
 {\cal O}\left(\frac{1}{\hat{l}^4}\right)
\end{eqnarray*}    
As noted earlier
\begin{eqnarray*}
\frac{1}{r_l} & = & \frac{Z_v}{\hat{l}(\hat{l}+1)} \\
              & \simeq & \frac{Z_v}{\hat{l}^2} 
\end{eqnarray*} 
This suggests that $E_{n,l}$ is a polynomial in $1/r_l$ with
constants which depend on the principal quantum number $n$.
Recalling eqn.(9), namely $\chi \sim E_s$ it is reasonable to
hypothesize a relationship of the form
\begin{eqnarray} 
\chi = \sum_{m=0} \frac{a_m(n)}{r_s^m}
\end{eqnarray} 
where m is an integer and the coefficients  $a_m(n)$ depend on the
principal quantum number. Eqn~(10) is a polynomial in $1/r_s$.  In
this spirit we  have attempted a  correlation between  the Pauling
electronegativity and our REEL co-ordinate ${\cal V} (\equiv
\sqrt{n-1}/r_s)$, namely
\begin{eqnarray} 
\chi(R_j) = \chi_0 + c_1{\cal V} (R_j) + c_2 {\cal V}^2(R_j)
\end{eqnarray} 
This quadratic  relationship is depicted  in fig. 5. The constants
employed were $\chi_0$ = 0.018, $c_1$ =  0.568 (a.u.)$^{-1}$ , and
$c_2$ = 0.117 (a.u.)$^{-2}$.     A  simple linear fit  was    also
attempted and  it is depicted by dashed  lines  in fig.~5. For the
linear  fit the  slope and intercept  are  1.101 (a.u.)$^{-1}$ and
-0.538 respectively.  The quadratic  fit correctly  reproduces the
observed parabolicity and has   a smaller standard  deviation.  It
should be preferred over the linear one.

	Given the  above  arguments the   relationship between the
Sanderson electronegativity  and our  REEL co-ordinate (fig.~2 and
eqn~(3))  follows in a natural fashion.  The Pauling and Sanderson
electronegativities are related as follows \cite{luco79}
\begin{eqnarray*}
\sqrt{\chi} = 0.21 SR  + 0.77
\end{eqnarray*} 
We find a similar linear relationship, but with the slope 0.19
(instead of 0.21) and intercept 0.79 (instead of 0.77). In any
case,  since $\chi$ is quadratic in $\cal V$ and in
{\it SR} it follows that 
\begin{eqnarray*}
 SR \propto \frac{\sqrt{n-1}}{r_s}
\end{eqnarray*} 

	The   above considerations  are   by  no means  a rigorous
derivation of  eqns.~(2) and (3).   They have simply motivated and
guided  our search  for  an  appropriate relationship between  the
electronegativity and the orbital radii. Based  on our findings we
claim that instead  of three variables  \{$\chi, r_s, n$\} one can
perhaps work  with the reduced  set \{$r_s,n$\} for most purposes.
The latter has  the advantage in the  sense that its  elements are
{\it non-empirical}. In  other words, borrowing a terminology from
the theory of critical  phenomena   in statistical  mechanics,  we
have    found that  the  electronegativity   is a {\it generalized
homogeneous function} of the principal quantum  number $n$ and the
orbital radius $r_s$.

	 Environments involving  \{Si,O,H\} have been suggested as
explanation      for     visible     PL     in    porous   silicon
\cite{bran92,tsai92,prok92}.   Oxygen,  fluorine    and    organic
radicals can get  introduced  into the  silicon system  during the
anodization of silicon in HF.  Substituted oxygen and fluorine may
give rise  to  $\nu_{Si-H}$ in  the  range (2050-2150  cm$^{-1}$).
Table  2  enumerates   some  of the  possible    environments with
frequencies calculated  using equation \ref{eq:3}.  The first four
entries of $\nu_{Si-H}$ fall  in the range 2050-2150  cm$^{-1}$ or
are  close to it.  The  next three entries pertain to environments
with  higher  $\nu_{Si-H}$.   The  larger  linewidths observed  in
porous silicon (20 cm$^{-1}$)   as opposed to the extremely  sharp
lines   (linewidth $\approx$ 1   \mbox{cm$^{-1}$}) in  crystalline
silicon suggest disorder. It is well known  that a large number of
dangling bonds are present in porous  silicon.  Hence we have also
considered dangling  bond environments.  The  last four entries in
table 2 suggest four such environments which may contribute to the
observed $\nu_{Si-H}$ line  (2050-2150 \mbox{cm$^{-1}$}) in porous
silicon.   Note   that the  substitution of   an   oxygen atom  in
(O$_3$)SiH  by a dangling   bond  reduces $\nu_{Si-H}$  from  2273
\mbox{cm$^{-1}$}  to  2092  \mbox{cm$^{-1}$}.Broad  lines are also
observed   at   $\approx$      850  \mbox{cm$^{-1}$}   and    1100
\mbox{cm$^{-1}$}.   These lines   have  been assigned to  Si-H$_2$
scissors mode   and Si-O-Si stretching  modes respectively  in the
past.  However, Si-F related frequencies \cite{yama83} also lie in
the range 870  \mbox{cm$^{-1}$} to 1030 \mbox{cm$^{-1}$}.  In view
of the  possible presence of fluorine   in porous silicon, caution
must   be  exercised in assigning   frequencies   to the  observed
infrared lines.

	In a recent work,  Schuppler and co-workers  \cite{schu95}
claim that   the visible PL in   porous silicon is due  to quantum
confinement effects  arising from  silicon particles   (not wires)
typically of size $< 1.5$ nm. Their claim which is based partially
on  IR  data is that   such  a cluster   is fully  passivated with
hydrogen. Being small,   the cluster would have  a  fair number of
dihydrides and trihydrides as  well as some monohydrides which are
predominantly  seen   in   larger clusters.     They  suggest that
calculations on passivated Si   clusters  with $\sim  10^2$  atoms
would elucidate the luminescence behavior of  PS. As we have noted
in   the  previous  paragraph,   $\nu_{Si-H}$  with dangling  bond
environments also  lie in the  observed range.  This suggests that
electronic   structure   calculations should  be   carried out for
clusters which are fully passivated as well as for dangling bonds.

	The past  assignments of the  vibrational spectra of PS to
specific transitions exhibit  some inconsistencies.  Our model has
limited value   in  making  exact  identifications,   but suggests
systematic variations  in  the   vibrational  spectrum with    the
addition (removal)  of other  elements  to  the environment.   For
example, Shanks and co-workers \cite{shan80}  state that the shift
of  the 2000 cm$^{-1}$ line on  annealing to 2100 cm$^{-1}$ is due
to the introduction  of  two dangling  bonds.  However, our   work
seems to indicate that the introduction of dangling bonds can only
lower the  vibrational frequency  as mentioned earlier.   Thus the
trends suggested  by the present work  can be used to improve upon
existing results.  Again,  our    model contradicts some   of  the
assignments due  to Gupta,  Colvin and George  \cite{gupt88} who
posit a decrease in the   vibrational frequency as the number  of
neighboring hydrogen atoms   are increased.  The  main utility  of
this  model thus  lies in  correlating changes  in the vibrational
spectrum  with specific changes  in the surface environment, which
is   known to   play     a   major   role   in   porous    silicon
photoluminescence.  The  PL   spectra   of  siloxene,  a   complex
considered responsible for the luminescence in porous silicon, can
also be varied by the  introduction of halides or organic radicals
\cite{bran92}.   In   porous  silicon,  post  anodization  surface
treatment  modifies the PL spectra  \cite{hou93}.  This is perhaps
due to the   formation of molecular   complexes on the  surface of
porous silicon.  Some insight into this  phenomena can be obtained
by  studying the   vibrational spectra.  Our   approach provides a
valuable tool to correlate the PL  spectra of various samples with
their vibrational spectra.   This can  serve to  characterize  the
porous silicon surface.

	 A     universal   relationship  between    the  Sanderson
electronegativity and our quantum scale \mbox{$\cal V$} defined by
eqn.~(\ref{eq:verma}) has been  established  which is seen  to  be
valid  for almost the entire  periodic table.  Since the Sanderson
and  Pauling   ($\chi  $) electronegativities   are  related, (ie,
$\sqrt{\chi}  \propto  SR)$    a  similar relationship    can   be
established  between the Pauling electronegativity and \mbox{$\cal
V$}.    This is depicted  in figure   5.   The infrared stretching
frequencies of  Si-H  and   Si-Cl  bonds  have been    related  to
\mbox{$\cal V$}.  The relationship does   not work  well when  the
system  is  highly electronegative, e.g.   SiCl$_4$.  This is also
evident   from the  saturation  effect   seen in  fig.2.   We  are
investigating  extensions of our  work to other defects in silicon
and other semiconductors as well as wagging and bending modes.  We
suggest  that assignments   of vibrational  spectra to  particular
defects in porous silicon must be done with caution.

\acknowledgements
	We  would like to acknowledge  support from  the Coucil of
Scientific and Industrial   Research, Government of India and  the
Department of Science and Technology,  Government of India. Useful
correspondence with Dr.A.Zunger is gratefully acknowledged.

\begin{thebibliography}{10}

\bibitem{canh90}
L.T.Canham, Appl. Phys. Lett. {\bf 57},  1046  (1990).

\bibitem{john95r}
G.~C. John and V.~A. Singh, Phys. Reports {\bf 263},  93  (1995).

\bibitem{bran92}
M.S.Brandt {\it et~al.}, Solid State Commun. {\bf 81},  307  (1992).

\bibitem{tsai92}
C.Tsai {\it et~al.}, Appl. Phys. Lett. {\bf 60},  1700  (1992).

\bibitem{prok94}
S.M.Prokes and O.J.Glembocki, Phys. Rev. B {\bf 49},  2238  (1994).

\bibitem{ohno92}
T.Ohno, K.Shiraishi, and T.Ogawa, Phys. Rev. Lett. {\bf 69},  2400  (1992).

\bibitem{smit59}
A.L.Smith and N.C.Angellotti, Spectrochemica Acta {\bf 15},  412  (1959).

\bibitem{luco79}
G.Lucovsky, Solid State Commun. {\bf 29},  571  (1979).

\bibitem{shi82}
T.S.Shi {\it et~al.}, Phys. Stat. Sol.(a) {\bf 74},  329  (1982).

\bibitem{sahu82}
S.N.Sahu {\it et~al.}, J. Chem. Phys. {\bf 77},  4330  (1982).

\bibitem{phil78}
J.C.Phillips, Comments in Sol. St. Phys. {\bf 9},  11  (1978).

\bibitem{zung80}
A.Zunger, Phys. Rev. B {\bf 22},  5839  (1980).

\bibitem{mied73}
A.R.Miedema, J. Less-Common Metals {\bf 32},  117  (1973).

\bibitem{sing82}
V.A.Singh and A.Zunger, Phys. Rev. B {\bf 25},  907  (1982).

\bibitem{vill88}
P.Villars and J.C.Phillips, Phys. Rev. B (RC) {\bf 37},  2345  (1988).

\bibitem{yeh92}
C.-Y. Yeh, Z.W.Lu, S.Froyen, and A.Zunger, Phys. Rev. B {\bf 45},  12130
  (1992).

\bibitem{john74}
{J. St.John} and A.N.Bloch, Phys. Rev. Lett. {\bf 33},  1095  (1974).

\bibitem{chel78}
J.R.Chelikowsky and J.C.Phillips, Phys. Rev. B {\bf 17},  2453  (1978).

\bibitem{burd82}
J.Burdett, Acc. Chem. Res. {\bf 15},  34  (1982).

\bibitem{zung78}
A.Zunger,  in {\em Structure and Bonding in Crystals}, edited by M.O.Keefe and
  A.Navrotsky (Academic Press, New York, 1978).

\bibitem{pear69}
W.B.Pearson,  in {\em Developments in the Structural Chemistry of Alloy
  Phases}, edited by B.C.Giessen (Plenum, New York, 1969).

\bibitem{shaw68}
R.W.Shaw, Phys. Rev. {\bf 174},  769  (1968).

\bibitem{sand60}
R.T.Sanderson, {\em Chemical Periodicity} (Reinhold Publishing Corp., New York,
  1960), pg. 16-56.

\bibitem{knif83}
N.Kniffler, B.Schroeder, and J.Geiger, J. Non-Cryst. Sol. {\bf 58},  153
  (1983).

\bibitem{wu83}
{Wu Shi-Qiang} {\it et~al.}, J. Non-Cryst. Sol. {\bf 59-60},  217  (1982).

\bibitem{unag80a}
T.Unagami, J. Electrochem. Soc. {\bf 127},  476  (1980).

\bibitem{unag80b}
T.Unagami, Jpn. J. Appl. Phys. {\bf 19},  231  (1980).

\bibitem{gupt88}
P.Gupta, V.Colvin, and S.M.George, Phys. Rev. B {\bf 37},  8234  (1988).

\bibitem{kato88}
Y.Kato, T.Ito, and A.Hiraki, Jpn. J. Appl. Phys. {\bf 27},  L1406  (1988).

\bibitem{koch93b}
F.Koch, V.Petrova-Koch, and T.Muschik, J.Lumin. {\bf 57},  271  (1993).

\bibitem{ogat95}
Y.Ogata, H.Niki, T.Sakka, and M.Iwasaki, J. Electrochem. Soc. {\bf 142},  195
  (1995).

\bibitem{dark53}
L.S.Darken and R.W.Gurry, {\em Physical Chemistry of Metals} (McGraw Hill, New
  York, 1953).

\bibitem{wabe63}
J.T.Waber {\it et~al.}, Trans. Mettall. Soc. AMIE {\bf 227},  717  (1963).

\bibitem{wats78}
R.E.Watson and L.H.Bennet, J. Phys. Chem. Solids {\bf 39},  1235  (1978).

\bibitem{simo73}
G.Simons and A.N.Bloch, Phys. Rev. B {\bf 7},  2754  (1973).

\bibitem{vill87}
P.Villars and F.Hulliger, J. Less-Common Metals {\bf 132},  289  (1987).

\bibitem{phil70}
J.C.Phillips, Rev. Mod. Phys. {\bf 42},  317  (1970); {J.A.Van Vechten},Phys.
  Rev. {\bf 182}, 891 (1969); {\bf 187}, 1007 (1969); Phys. Rev. B {\bf 7},
  1479 (1973).

\bibitem{prok92}
S.M.Prokes {\it et~al.}, Phys. Rev. B {\bf 45},  13788  (1992).

\bibitem{yama83}
K.Yamamoto, T.Nakanishi, H.Kasahara, and K.Abe, J. Non-Cryst. Sol. {\bf 59-60},
   213  (1983).

\bibitem{hou93}
X.Y.Hou {\it et~al.}, Appl. Phys. Lett. {\bf 62},  1097  (1993).

\bibitem{shan80}
H.Shanks {\it et~al.}, Phys. Stat. Sol. (b) {\bf 100},  43  (1980).

\bibitem{schu95}
S.Schuppler {\it et~al.}, Phys. Rev. B {\bf 52},  4910  (1995).

\bibitem{read92}
A.J.Read {\it et~al.}, Phys. Rev. Lett. {\bf 69},  1232  (1992).

\bibitem{sand92}
G.D.Sanders and Y-C.Chang, Appl. Phys. Lett. {\bf 60},  2525  (1992).

\end{thebibliography}

\begin{figure}
\caption{Row-wise plots of the Sanderson electronegativity ${\it
SR}(R_j)$ and $1/r_s(R_j)$,  where $r_s(R_j)$ is the $l=0$ orbital
radius of the element. The principal quantum number of the valence
shell is denoted by n.}
\end{figure}
\begin{figure}
\caption{Linear relationship between the Sanderson electronegativity
$SR(R_j)$       and     the      quantum    scale   ${\protect\cal
V}(R_j)=\protect\sqrt{n(R_j)-1}/r_s(R_j)$   where    $n(R_j)$  and
$r_s(R_j)$  denote the  principal  quantum number  of  the valence
shell and the $l=0$ orbital radius of the element respectively.}
\end{figure}
\begin{figure}
\caption{Linear relationship between the Si-H band stretching
frequency     $\nu_{Si-H}$   and         the     quantum     scale
 {$\protect\displaystyle\sum_   {j=1}^4$}\mbox{$\cal    V$}$(R_j)$
 where   the  summation runs over  the  four  bonds of the central
 silicon atom.}
\end{figure}
\begin{figure}
\caption{Linear relationship between the Si-Cl bond stretching
frequency $\nu_{Si-Cl}$ and the quantum scale
{$\protect\displaystyle\sum_{j=1}^4$}\mbox{$\cal V$}$(R_j).$}  
\end{figure}
\begin{figure}
\caption{Plot of the Pauling electronegativity $\chi$ vs the
quantum scale ${\protect\cal V}$. The solid (dashed) line is a
quadratic (linear) fit to the data.}
\end{figure}
\begin{table}
\caption{Stretching mode Si-H frequencies ($\nu_{Si-H}$) for well
established environments in crystalline  (c-Si) and  amorphous (a-Si)
silicon.   Data   for   c-Si  are  taken  from   Shi  {\it   et  al.}
\protect\cite{shi82}  and  for  a-Si  from  Kniffler   {\it  et  al.}
\protect\cite{knif83}. The symbol `d' represents a dangling bond.}
\begin{tabular}{|c|c|c|c|}
No. & \hspace*{0.4in}System\hspace*{0.4in} &\hspace*{0.3in}
$\nu_{Si-H}$ (\mbox{cm$^{-1}$})\hspace*{0.3in} &
\hspace*{0.3in}Remarks \hspace*{0.3in}\\ \hline 
 1 & (SiSiSi)SiH & 1980-1990 & c-Si\\
2 & (CSiSi)SiH & 2028-2030 & c-Si\\
3 & (SiSi)SiH$_2$ & 2055-2066 & c-Si\\
4 & (CCSi)SiH & 2083 & c-Si\\
5 & (CSi)SiH$_2$ & 2105-2107 & c-Si\\
6 & (OSi)SiH$_2$ & 2160-2162 & c-Si\\
7 & (OC)SiH$_2$ & 2210-2218 & c-Si\\
8 & (O$_2$Si)SiH & 2193 & a-Si\\
9 & (O$_3$)SiH & 2247 & a-Si\\
10 & (O$_2$)SiH$_2$ & 2219 & a-Si\\
11 & (OSi$_2$)SiH & 2100 & a-Si\\
12 & (Sidd)SiH & 1830-1840 & c-Si\\
13 & (CSid)SiH & 1925-1931 & c-Si\\
14 & (Sid)SiH$_2$ & 1957-1965 & c-Si\\ 
\end{tabular}
\end{table}

\begin{table}
\caption{Stretching mode Si-H frequencies ($\nu_{Si-H}$) for possible
environments in porous silicon. The $\nu_{Si-H}$ have been calculated
on the basis of eqn.~(\protect\ref{eq:3}). The symbol `d' represents a
dangling bond.} 
\begin{tabular}{|c|c|c|} 
No. & System & $\nu_{Si-H}$ (\mbox{cm$^{-1}$})
 \\ \hline
   & \hspace*{2in} & \hspace*{2in} \\
1 & (Si)SiH$_3$ & 2106 \\
2 & (OSi)SiH$_2$ & 2146 \\
3 & (OC)SiH$_2$ & 2175 \\
4 & (SiO$_2$)SiH & 2186 \\
5 & (F)SiH$_3$ & 2223 \\
6 & (O$_3$)SiH & 2273 \\
7 & (F$_2$)SiH$_2$ & 2293 \\
8 & (Od)SiH$_2$ & 2052 \\
9 & (Fd)SiH$_2$ & 2082 \\
10 & (O$_2$d)SiH & 2092 \\
11 & (F$_2$d)SiH & 2152 \\
\end{tabular}

\end{table} 

\end{document}